\pdfoutput=1

\documentclass{iau}
\usepackage{graphicx}
\newcommand\arcsec{\mbox{$^{\prime\prime}$}}%
\newcommand\arcmin{\mbox{$^\prime$}}%

\title[Feedback from The Nuclear Activities] 
{A New Perspective on the Radio Active Zone at The Galactic
       Center \\ -- Feedback from  Nuclear Activities}

\author[Jun-Hui Zhao, Mark R. Morris \& W. M. Goss]   
{Jun-Hui Zhao$^1$, Mark R. Morris$^2$
 \and W. M. Goss$^3$}

\affiliation{$^1$Harvard-Smithsonian CfA, 60
Garden Street, MS 78, Cambridge, MA 02138 \\ email: {\tt jzhao@cfa.harvard.edu} \\[\affilskip]
$^2$Department of Physics and Astronomy, University of California, Los Angeles, CA 90095  \\email: {\tt morris@astro.ucla.edu}
\\[\affilskip]
$^3$NRAO, P.O. Box O, Socorro, NM 87801, USA \\email: {\tt mgoss@aoc.nrao.edu}
}

\pubyear{2013}
\volume{303}  
\pagerange{119--126}
\jname{The Galactic Center}
\editors{A.C. Editor, B.D. Editor \& C.E. Editor, eds.}
\begin{document}

\maketitle

\begin{abstract}
Based on our  deep image of Sgr A using broadband data observed 
with the Jansky VLA\footnote{The Jansky Very Large
Array (JVLA) is operated by the National Radio Astronomy Observatory
(NRAO). The NRAO is a facility of the National Science Foundation
operated under cooperative agreement by Associated Universities, Inc.} 
at 6 cm, we present a new perspective of the radio bright zone at 
the Galactic center. We further show the radio detection of the 
X-ray Cannonball, a candidate neutron star associated with the 
Galactic center SNR Sgr A East. The radio image is compared with 
the Chandra X-ray image to show the detailed structure of the 
radio counterparts of the bipolar X-ray lobes. The bipolar lobes 
are likely produced by the winds from the activities within Sgr A West, 
which could be collimated by the inertia of gas in the CND, 
or by the momentum driving of Sgr A*; and the poloidal magnetic 
fields likely play an important role in the collimation. The 
less-collimated SE lobe, in comparison to the NW one, is perhaps 
due to the fact that the Sgr A East SN might have locally reconfigured 
the magnetic field toward negative galactic latitudes. In agreement 
with the X-ray observations, the time-scale of $\sim1\times10^4$ yr 
estimated for the outermost radio ring appears to be comparable to 
the inferred age of the Sgr A East SNR.  
\end{abstract}

\firstsection 
\section{Introduction}
In the inner 40 parsecs, from the 1 Msec integration Chandra X-ray image,
\cite{morr03} showed  remarkable X-ray bipolar lobe structures 
($\sim$10 parsecs) centered at Sgr A*, extending in the direction 
parpendicular to the Galactic plane. The authors pointed out that 
a number of emission clumps in the bipolar lobes suggests a series 
of ejections from the circumnuclear disk (CND) or Sgr A West 
occurred on time scales of hundreds to thousands of years. The X-ray 
bipolar lobes imply ongoing activity from the circumnuclear 
region surrounding Sgr A*. Within this region, a well known radio 
emission shell, Sgr A East, is interpreted simply as a SNR (\cite[Jones 
1974, Green 1984, Goss et al. 1985]{jone74,gree84,goss85}); the 
age of Sgr A East inferred from the various models spans a large 
range, from 1700 yr to 5$\times10^4$ yr. Using X-ray data observed 
with the Chandra Observatory, \cite{park05} found a hard compact 
X-ray source, CXOGC J174545.5-285829, having unusual X-ray 
characteristics compared to most other X-ray sources near the 
Galactic center. The authors suggested that the X-ray object could 
be identified as a high-velocity neutron star, produced from the 
core-collapse supernova (SN) explosion that created the Galactic 
center supernova remnant (SNR), Sgr A East.

\begin{figure}[b]
\begin{center}
\includegraphics[width=5.25in,angle=0.0]{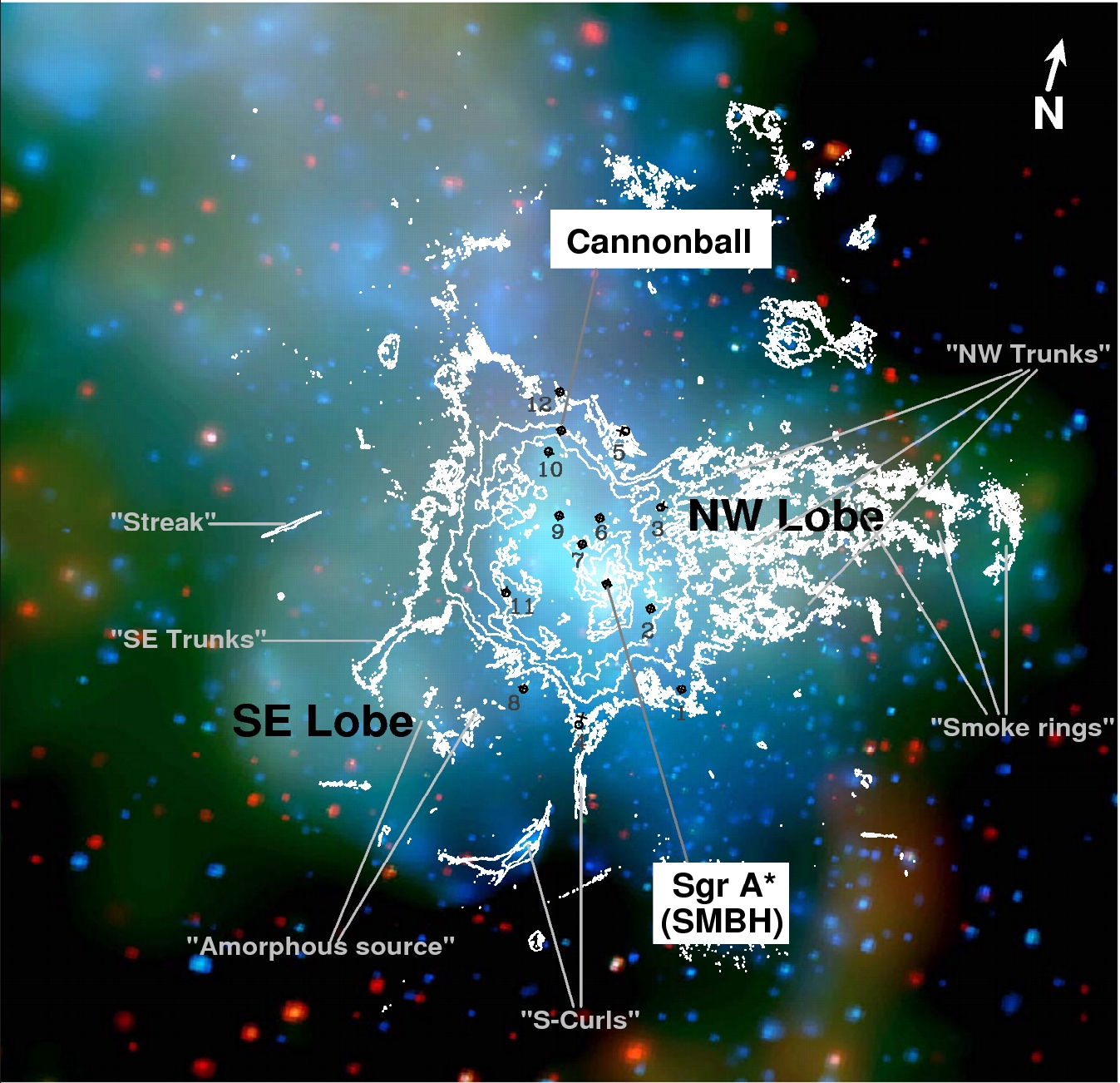}
\caption{The JVLA image at 5.5 GHz (contours, 
$\theta_{\rm FWHM}$=1.6\arcsec$\times$0.6\arcsec) overlaid on 
the Chandra X-ray image (background  
presented online by \cite[Marcoff 2010]{mark10}). The fourteen 
compact X-ray sources (open circles) and 
their radio identifications (crosses) are marked 
(see \cite[Zhao, et al. 2013]{zhao13}). 
}
\label{fig1}
\end{center}
\end{figure}

\section{Radio bright zone}
\begin{figure}[b]
\vspace*{-1.25 cm}
\begin{center}
\includegraphics[width=5.6in,angle=0.0]{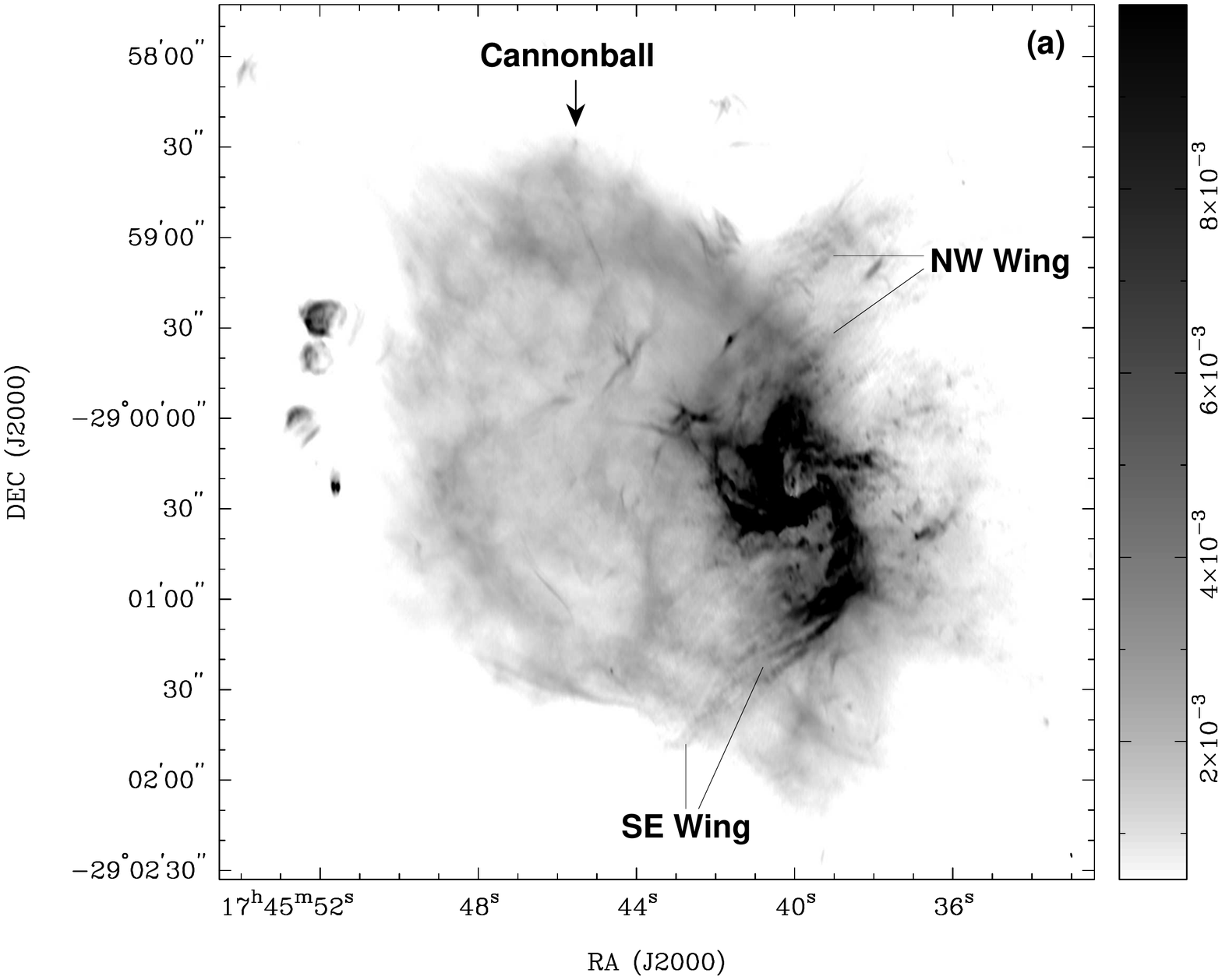}
\vspace*{-1.0 cm} \\
\includegraphics[width=2.6in,angle=0.0]{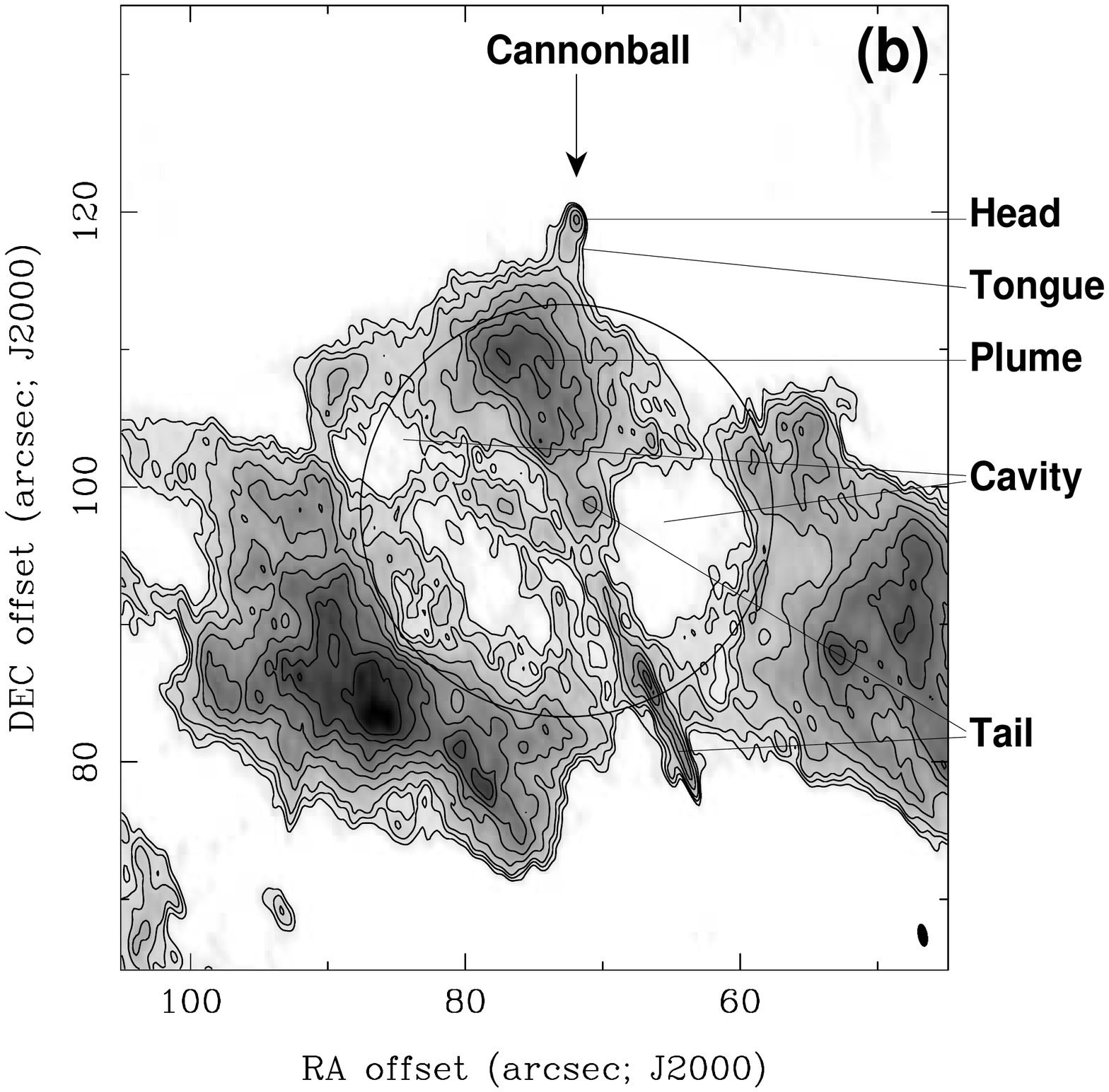}
\includegraphics[width=2.6in,angle=0.0]{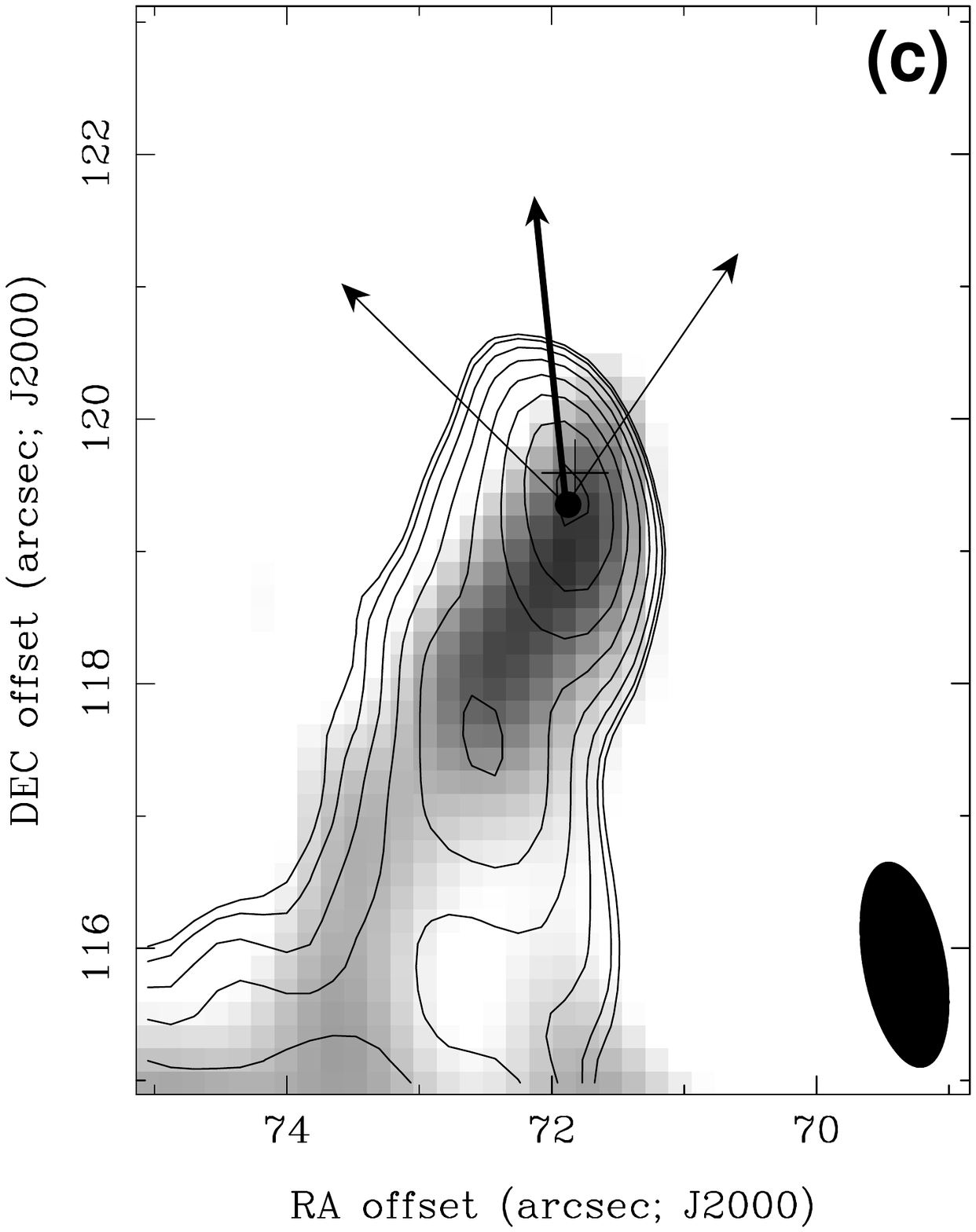}
\caption{
(a) A new 5.5-GHz JVLA image  of Sgr A East and West along with the radio 
counterpart of the Cannonball.
(b) The radio structure  of
the PWN; the radio components possibly associated with
the PWN are labelled.
(c) Proper motion vector (bold arrow) of the Cannonball 
with 3$\sigma$ uncertainty in direction (light arrows) 
anchored at the peak of the radio source observed in 2012 
(contours) that is overlaid on the JVLA
image observed in 1987 (grey scale). The FWHM beam sizes of above images are $\theta_{\rm FWHM}$=1.6\arcsec$\times$0.6\arcsec.
}
\label{fig2}
\end{center}
\end{figure}

We have observed Sgr A with the JVLA in the B \& C arrays 
using the broadband (2 GHz) continuum mode at 5.5 GHz covering the 
central 13\arcmin\, (30 pc) region of the radio bright zone at the 
Galactic center. With  
MS-MFS clean algorithm of CASA, we constructed a deep image, 
achieving an rms noise level of 0.01 mJy beam$^{-1}$, or a dynamic 
range 80,000:1. The data reduction procedure is described 
by Zhao et al. (2013 in preparation).  Our observations 
have revealed the detailed structures of both previously known and 
newly identified radio sources in this region. Numerous compact radio 
sources have been detected at a level of $\sim$0.1 mJy beam$^{-1}$. 
In general, the emission structure of the radio bright zone at 5 GHz
is characterized by Sgr A West (the HII minispiral), Sgr A East
(SNR) and radio filaments (\cite[Morris et al. 2013]{morr13}).
The deep radio image has been compared with the Chandra X-ray
(Fig.\,\ref{fig1}) and HST/NICMOS Paschen-$\alpha$ images 
(\cite[Wang et al. 2010]{wang10}). In the
radio, broad ``Wings'' (Fig.\,\ref{fig2}a) extend 100\arcsec\,(4 pc) 
from the tips of the minispiral (Sgr A West) to 
the NW and SE; the NW Wing appears to be along with several radio emission 
``Trunks'' forming an elongated radio lobe with a size of 
6\arcmin$\times$3\arcmin\,(14$\times$7 pc) and oriented perpendicular 
to the Galactic plane, in projection. In the outer region of the NW lobe, 
a progression of three emission rings (``Smoke rings'') is present, 
indicating a thermal free-free origin 
based on an almost identical structure in the Paschen-$\alpha$ image.
The NW radio lobe matches well with its X-ray counterpart (Fig.\,\ref{fig1}).
An amorphous radio emission structure at the tip of the SE Wing and an
emission trunk (120\arcsec$\times$20\arcsec, or 4.6$\times$0.8 pc)
are located in the SE X-ray lobe that appears to be confined by
the ``Streak'' and the southern ``Curl'' filaments  
(\cite[Morris et al. 2013]{morr13}). Since the implied
accretion rate onto the central SMBH is 
far below the Eddington limit, the direct activity from accretion 
appears to contribute little to the feedback to the ISM in this region
at the present time. The bipolar X-ray lobes are likely produced by the winds from the 
activities within Sgr A West, which could be collimated by the inertia of 
the gas in the CND or by the momentum driving of Sgr A*; and 
the poloidal magnetic fields likely play an important role in the collimation.  
The less-collimated SE lobe, in comparison to the NW  one, is perhaps 
due to the fact that the Sgr A East SN might have locally reconfigured 
the magnetic field toward negative galactic latitudes. The  massive
stars near the ``Smoke rings'' (personal communication from 
Hui Dong \& Jon Mauerhan, 2013) are likely the main sources for  
the ionization of the local gas.
If the filamentary structure perpendicular to the major-axis of 
the NW lobe is a wave pattern in a magnetized-rotating plasma, 
propagating from Sgr A West with an Alfv$\acute{\rm e}$n
velocity $v_A\approx 218~{\rm km~s^{-1}}
\left[B\over 1~{\rm mG}\right]
\left[n_i\over 10^2~{\rm cm}^{-3}\right]^{-1/2}$,
a time scale of $\sim1\times10^4$ yr is estimated 
for the wave traveling to the outermost ring, 
given a B-field stength of a few mG 
and an ion number density of $\sim10^2$ cm$^{-3}$  
in the nuclear environment.
In good agreement with the estimate for the X-ray lobes, 
this time scale, by coincidence,
appears to be comparable to the age of the Sgr A East SNR inferred from the
Cannonball (\cite[Zhao et al. 2013]{zhao13}).

\section{Radio counterpart of the Cannonball}

The radio object was detected both in the JVLA image from 2012 observations 
at 5.5 GHz and in archival VLA images from observations in 1987 at 
4.75 GHz and in the period from 1990 to 2002 at 8.31 GHz. The radio 
morphology of this object is characterized as a compact, partially 
resolved point source located at the northern tip of a radio ``tongue'' 
(Fig.\,\ref{fig2}b) similar to the X-ray structure observed by Chandra 
(\cite[Park et al. 2005]{park05}). Behind the Cannonball, a radio counterpart to the X-ray 
plume is observed. This object consists of a broad radio plume with 
a size of 30\arcsec$\times$15\arcsec, followed by a linear tail 
having a length of 30\arcsec. The compact head and broad plume 
sources appear to have relatively flat spectra ($\propto\nu^\alpha$) 
with mean values of $\alpha = -0.44\pm 0.08$ and $-0.10\pm0.02$, 
respectively, and the linear tail shows a steep spectrum with 
the mean value for $\alpha$ of $-1.94\pm 0.05$. The total radio luminosity 
integrated from these components is $\sim8\times10^{33}$ 
erg s$^{-1}$, while the emission from the head and tongue 
amounts to only $\sim1.5\times 10^{31}$ erg s$^{-1}$. Based 
on the images obtained from the two epochs' observations at 
5 GHz, we infer a proper motion of the object: 
$\mu_\alpha = 0.001 \pm 0.003$ arcsec yr$^{-1}$ and 
$\mu_\delta = 0.013\pm 0.003$ arcsec yr$^{-1}$ 
(see Fig.\,\ref{fig2}c). With an implied velocity of 500 
km s$^{-1}$, a plausible model can be constructed in 
which a runaway neutron star surrounded by a pulsar wind nebula 
(PWN) was created in the event that produced Sgr A East. 
The inferred age of this object, assuming that its origin 
coincides with the center of Sgr A East, is approximately 9000 yr.

\end{document}